%% file: ms.tex
\documentclass{emulateapj}

\newcommand{\kms}{km\,s$^{-1}$}     
  
\newcommand{\lya}{Lyman-$\alpha$}

\newcommand{\hi}{\ion{H}{1}}    
\newcommand{\os}{\ion{O}{6}}   \newcommand{\cf}{\ion{C}{4}} 
\newcommand{\sif}{\ion{Si}{4}} \newcommand{\nf}{\ion{N}{5}}
   \newcommand{\sit}{\ion{Si}{3}}
\newcommand{\ct}{\ion{C}{3}}   \newcommand{\siw}{\ion{Si}{2}}
\newcommand{\cw}{\ion{C}{2}}   \newcommand{\oi}{\ion{O}{1}}
\newcommand{\few}{\ion{Fe}{2}} \newcommand{\alw}{\ion{Al}{2}}
\newcommand{\cts}{\ion{C}{2}$^*$} \newcommand{\alth}{\ion{Al}{3}}
\newcommand{\feth}{\ion{Fe}{3}} 
\newcommand{\tm}{\tablenotemark} 
\newcommand{\tn}{\tablenotetext}

\begin{document}

\title{Multiphase plasma in Sub-Damped Lyman-Alpha Systems: A Hidden 
  Metal Reservoir\footnote{Based on observations taken under programme 
  IDs 072.A-0442(A) and 078.A-0164(A) with the Ultraviolet and Visual
  Echelle Spectrograph (UVES) on the Very Large Telescope (VLT) Unit 2
  (Kueyen) at Paranal, Chile, operated by ESO.}}

\author{Andrew J. Fox\altaffilmark{1}, Patrick Petitjean\altaffilmark{1,2},
 C\'edric Ledoux\altaffilmark{3}, And Raghunathan Srianand\altaffilmark{4}}
\altaffiltext{1}{Institut d'Astrophysique de Paris, 
  UMR 7095 CNRS, Universit\'e Pierre et Marie Curie, 
  98bis, Boulevard Arago, 75014 Paris, France; fox@iap.fr}
\altaffiltext{2}{LERMA, Observatoire de Paris,
  61 Avenue de l'Observatoire, 75014 Paris, France; ppetitje@iap.fr}
\altaffiltext{3}{European Southern Observatory, Alonso de C\'ordova
  3107, Casilla 19001, Vitacura, Santiago 19, Chile; cledoux@eso.org}
\altaffiltext{4}{IUCAA, Post Bag 4, Ganesh Khind, Pune 411 007, India;
 anand@iucaa.ernet.in}
\shorttitle{Fox et al.}
\shortauthors{Plasma in a Sub-DLA} 

\begin{abstract}
We present a VLT/UVES spectrum of a proximate sub-damped \lya\
(sub-DLA) system at $z_{\rm abs}=2.65618$ toward the quasar Q0331-4505
($z_{\rm qso}=2.6785\!\pm\!0.0030$).
Absorption lines of \oi, \siw, \sit, \sif, \cw, \ct, \cf, \few, \alw, and \os\
are seen in the sub-DLA, which has a neutral hydrogen column
density log\,$N_{\rm H\,I}=19.82\pm0.05$. The absorber is at a velocity of
$1\,820\pm250$\,\kms\ from the quasar; however, its low
metallicity [O/H]=$-1.64\pm0.07$, lack of partial coverage, lack of
temporal variations between observations taken in 2003 and 2006, and
non-detection of \nf\ imply the absorber is not a genuine intrinsic system. 
By measuring the \os\ column density and assuming equal metallicities
in the neutral and ionized gas, we determine the column density of hot
ionized hydrogen in this sub-DLA, and in two other sub-DLAs with \os\ drawn
from the literature. Coupling this with determinations of
the typical amount of warm ionized hydrogen in sub-DLAs, 
we confirm that sub-DLAs are a more important metal
reservoir than DLAs, in total comprising at least 
6--22\% of the metal budget at $z\approx2.5$. 
\end{abstract}

\section{Introduction}
Highly ionized gas is ubiquitous in galactic halos at high redshift. 
This is known from the detection of \cf\ and \os\ absorption in
damped \lya\ \citep[DLA, i.e. systems with 
log\,$N_{\rm H\,I}\!>\!20.3$;][]{Le98, WP00, Fo07a, Fo07b} 
and sub-DLAs \citep[systems with 19.0$<$log\,$N_{\rm
    H\,I}\!<\!20.3$;][]{DZ03, Pe03, Pe07}.  
These studies have found that \cf\ is present in every DLA where data
exists, and that \os\ absorption, tracing a hotter phase of gas, is
present in at least 40\% of DLAs. Detections of \cf\ and \os\
absorption have also been reported \citep{Be94, KT97, KT99} in Lyman
Limit Systems (LLSs), a lower \hi\ column density class of QSO
absorber that may be tracing the extended outer regions of 
galactic halos. 

Physically, there are two principal processes that are expected to
generate hot plasma in galactic halos. The first is star formation,
which leads to supernovae and the subsequent production of hot,
shock-heated interstellar gas, possibly in the form of winds
\citep{OD06, KR07, Fa07}. The second is accretion, which results in
the shock-heating of infalling intergalactic gas to temperatures of
10$^{5-7}$\,K \citep{Da01, FB03, Ka05}. Given the low metallicities and low
densities expected in these hot halo environments, the cooling times
will be very long ($\gtrsim10^{10}$\,yr for gas at $T=10^6$\,K, 0.01 solar
metallicity, and $n=10^{-3}$\,cm$^{-3}$), enabling metals to become
locked up. Indeed, approximately half of the metals produced in stars
by a redshift of two are yet to be observed \citep{Bo05, Bo06, Bo07,
  SF07}, and hot halos represent a potential harbor for these metals.

It has recently become apparent that sub-damped \lya\ systems
(sub-DLAs, with $19.0\!<\!{\rm log}\,N_{\rm H\,I}\!<\!20.3$, also known as a
Super-LLSs) play a significant role in the metal budget 
\citep{Pe05, Pe07, Pr06, Ku07}, but
the contribution of the \os-phase in sub-DLAs has
not been studied before; indeed, we are aware of only one other
published detection of \os\ in a sub-DLA \citep{Si02}.
We present in this paper
observations and ionization modelling of a sub-DLA showing
a detection of \os.

\section{Observations}
The spectra of QSO \object{Q0331-4505} were taken with the
Ultraviolet-Visual Echelle Spectrograph \citep[UVES;][]{De00} at the
8.2m Very Large Telescope Unit 2 (VLT/UT2) at Paranal, Chile during
observing runs in 2003 and 2006. The data reduction was performed
using the UVES pipeline described in \citet{Ba00}.
The spectra were co-added in wavelength regions covered in both runs,
and have a spectral resolution (FWHM) of 6.6\,\kms\ ($R$=45\,000). 

A strong \lya\ absorber exists in the spectrum at $z_{\rm abs}=2.65618$.
This redshift is measured using the velocity of the strongest component of
absorption in \ion{Si}{2} $\lambda$1304, and defines the
velocity zero-point in the following discussion. 
Using a Voigt profile fit to the damping wings of the \lya\ line, we find 
log\,$N_{\rm H\,I}=19.82\pm0.05$, identifying the absorber as a sub-DLA. 
In Figure 1 we show the absorption line profiles of \oi, \cw, \siw, 
\few, \alw, \feth, \ct, \sit, \sif, \cf, and \os, which all show
detections in the sub-DLA. No \cts, \alth, or \nf\ absorption is seen. 
Both lines of the \os\ doublet display identical optical
depth profiles over a velocity range of 175\,\kms, verifying the
genuine detection of \os\ absorption, free from blending with the
\lya\ forest. Voigt profile fits were conducted
using the VPFIT software package\footnote
  {Available at http://www.ast.cam.ac.uk/$\sim$rfc/vpfit.html}, 
and are included in Figure 1.
The fits to each ion were completely independent.
A summary of the key properties of the sub-DLA is given in Table 1.
\input{tab1.tex}

\begin{figure*}
\epsscale{1.2}
\plotone{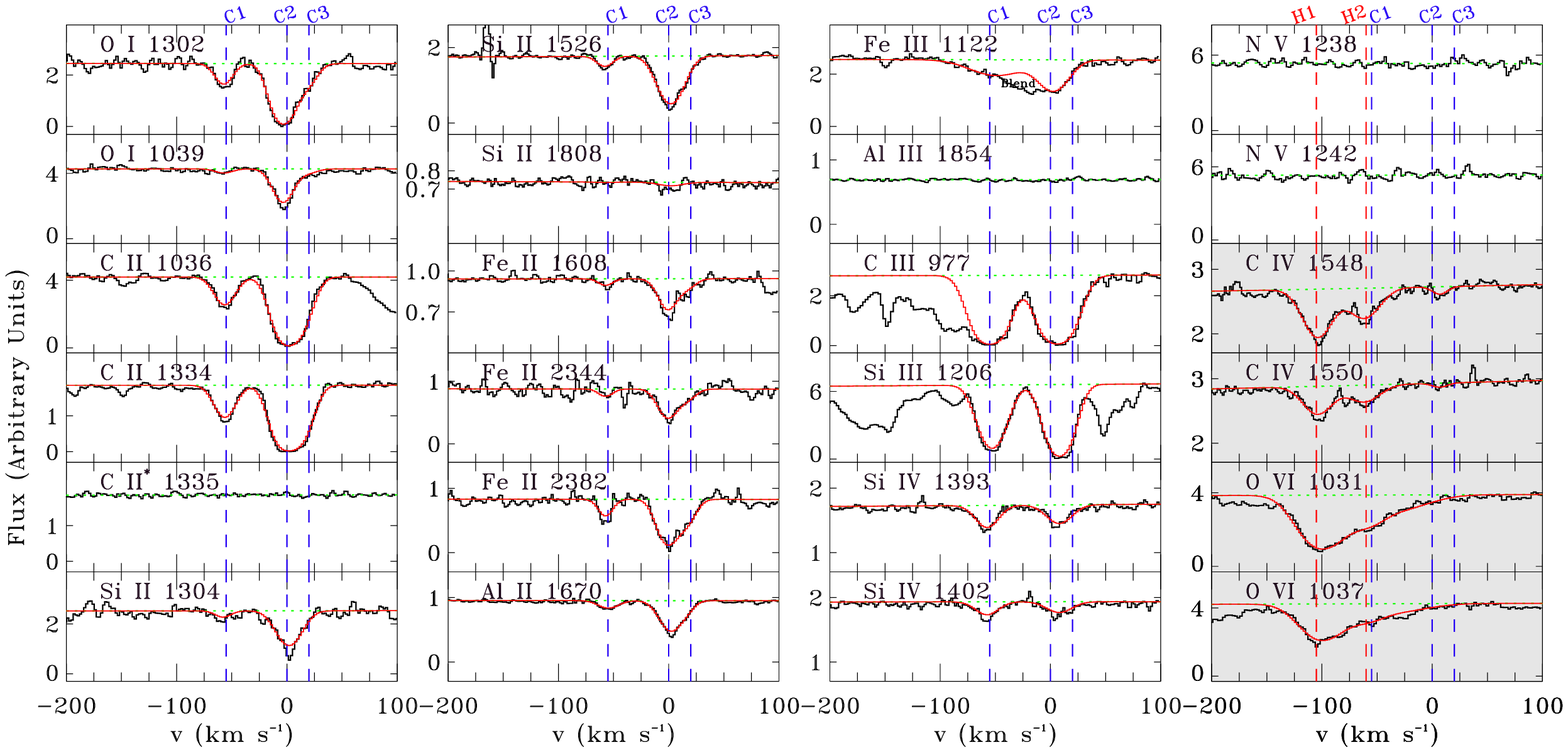}
\caption{VLT/UVES spectra of neutral, low-ion, and high-ion absorption
  in the sub-DLA at $z_{\rm abs}=2.65618$ toward Q0331-4505.
  This redshift, where the strongest component of low-ion
  absorption falls, defines the zero-point of the velocity scale. 
  Our overall Voigt component fits to each profile are shown with
  solid red lines. Vertical dashed lines illustrate the positions of the 
  three low-ion components (C1, C2, and C3) and two high-ion
  components (H1 and H2).
  Note that in some panels, the profiles are shown over a limited
  y-range, for clarity.
  We have shaded the \cf\ and \os\ panels to highlight the different
  structure in the high-ion absorption profiles.}  
\end{figure*}

We identify three principal components of low-ion absorption, 
at $-$55, 0, and 20\,\kms, which we name C1, C2, and C3, 
respectively. C2 is the strongest component. C3 is implied by the need to
fit the positive-velocity wing of the low-ion lines, but its
column density is relatively low, so we ignore it in the following
discussion. C1 and C2 are all seen in \oi, \siw, \alw, \few, \cw, \ct,
\sit, and \sif, and \cf.
We also identify two components of high-ion absorption seen in 
\os\ and \cf, at $-$100 and $-$60\,\kms, which we designate H1 and H2,
respectively. The \cf\ seen near $-$60\,\kms\ may have
contributions from both H2 and C1.

\section{Proximity to the Quasar}
Proximate absorbers separated from the background quasar by
d$v\!<$5\,000\,\kms\ [sometimes d$v\!<$3\,000\,\kms, where 
d$v$=$c|z_{\rm qso}\!-\!z_{\rm abs}|/(1\!+\!z_{\rm abs})$] are 
often removed from samples of intervening absorbers because they may
be either ejected or ionized (or both) by the QSO itself
\citep{We79, Fo86}.  
We measured the observed wavelengths of the \ion{N}{5}, \ion{O}{1}, 
\cf, \ion{He}{2}, and \ion{C}{3}] QSO emission lines in the
\object{Q0331-4505} spectrum, and derived the QSO redshift by
accounting for the average velocities of these lines (relative to the
systemic redshift) as listed in \citet{TF92}.
This yielded $z_{\rm qso}=2.6785\!\pm\!0.0030$\footnote{Improving
  the estimate of $z_{\rm qso}$=2.6 by \citet{Ma93}.},
implying the sub-DLA is separated from the quasar by a velocity of 
d$v=1\,820\!\pm\!250$\,\kms. 
Assuming a WMAP 3-year cosmology 
\citep[$H_0$=73\,\kms\,Mpc$^{-1}$, $\Omega_{\rm M}$=0.24, 
$\Omega_{\Lambda}$=0.76;][]{Sp07},
we use the relation given in \citet{Ph02} to find that 
the Hubble parameter at $z=2.65$ is $H(2.65)$=257\,\kms\,Mpc$^{-1}$. 
Therefore, assuming the sub-DLA motion is 
purely due to Hubble flow (i.e., ignoring peculiar velocities), we
find a separation of $\approx7.1$\,Mpc from the quasar. 

Despite the proximity to the quasar, 
several lines of evidence suggest this sub-DLA is \emph{not}
an intrinsic system arising near the central engine of the AGN: 
(1) the metallicity is too low \citep{Pe94};
(2) \nf, which is expected in intrinsic systems \citep{Ha97}, is not
detected;  
(3) no excited states (e.g. \ion{C}{2}$^*$ or \ion{Si}{2}$^*$) are
seen in absorption, as 
would be detected in intrinsic absorbers \citep[e.g.][]{Sr00}; 
(4) there is no evidence of partial coverage of the background source,
which is found in intrinsic absorbers \citep{Ar05};
(5) there is no evidence for temporal variability between our two
observation epochs in 2003 and 2006.

\section{Metallicity and Photoionization Modelling}
We use the \oi/\hi\ ratio to determine the metallicity of the
gas. Under interstellar conditions, \oi\ is thought to follow \hi\
closely due to charge exchange reactions \citep{FS71}, 
implying that the \oi/\hi\ ratio accurately determines
O/H without having to apply ionization corrections.
Using a Voigt profile fit to \oi\ $\lambda$1302 and  
$\lambda$1039, we find log\,$N$(\oi)=14.84$\pm$0.05 
(total over the three components). 
Coupling this with log\,$N$(\hi)=19.82$\pm$0.05, we find
[\oi/\hi]=[O/H]=$-$1.64$\pm$0.07\footnote{Throughout this paper we use
  the solar oxygen abundance (O/H)$_\odot=10^{-3.34}$ from
  \citet{As04}.}

To determine the ionization balance in the photoionized gas,
we ran a grid of CLOUDY \citep[version C06.02, described in][]{Fe98} 
photoionization models to component C2 at different 
ionization parameter $U\equiv n_{\gamma}/n_{\rm H}$.
The models assume the gas exists in a plane-parallel, uniform slab, and take
the metallicity [O/H] and $N_{\rm HI}$ as fixed inputs.
We assume the fraction of the total \hi\ column that exists in C2 is
the same as the fraction of the total \oi\ column in C2, which is 86\%.
To specify the radiation field, we take an unattenuated quasar
(QSO) spectrum, i.e. a power-law with the form 
$F_\nu\propto\nu^{-1.5}$ at $\lambda\!<$1216\,\AA\ \citep{HM96}
normalized by the measured QSO
$V$-band magnitude of 17.9 and $z_{\rm qso}=2.6785$, which yield a total
QSO luminosity at the Lyman Limit 
$L_{912}$=$1.2\times10^{31}$\,erg\,s$^{-1}$\,Hz$^{-1}$
when extrapolating the $V$-band flux as described in \citet{Gu07}. 
Making use of the QSO-absorber luminosity distance of 7.1\,Mpc, we
then find that the flux of ionizing radiation from the QSO at the
absorber is 
$F_{912}$=$2.0\!\times\!10^{-21}$\,erg\,cm$^{-2}$\,s$^{-1}$\,Hz$^{-1}$, 
\emph{lower} than the extragalactic background (EGB) at $z$=2.65, which
has $F_{912}$=$9.6\!\times\!10^{-21}$\,erg\,cm$^{-2}$\,s$^{-1}$\,Hz$^{-1}$
\citep{HM96}.
The QSO radiation field contains more high-energy photons
than the EGB, boosting the production of \sif\ and \cf, so the
radiation field we use in the CLOUDY modelling includes both the EGB
and the normalized QSO contribution.

We found the value of log\,$U$ that matches the \sif/\siw\ column
density ratio measured in C2. This value, log\,$U$=$-$3.32$\pm$0.10, 
also matches the observed \oi\ column density. We then found that 
[C/O]=$-$0.3 was necessary to reproduce the column densities of \cw\
and (within a 2$\sigma$ error) \cf. Finally we checked that the model
is consistent with the \ct\ and \sit\ lower limits. This model is
shown in Figure 2 and is characterized by a hydrogen ionization
fraction $x_{\rm H}$=0.64, 
and an ionized hydrogen column density log\,$N_{\rm H\,II}$=20.00. 
We assume the same conditions apply in C1 and C3, giving a total 
log\,$N_{\rm Warm\,H\,II}$=20.06. 

\begin{figure}
\epsscale{1.15}
\plotone{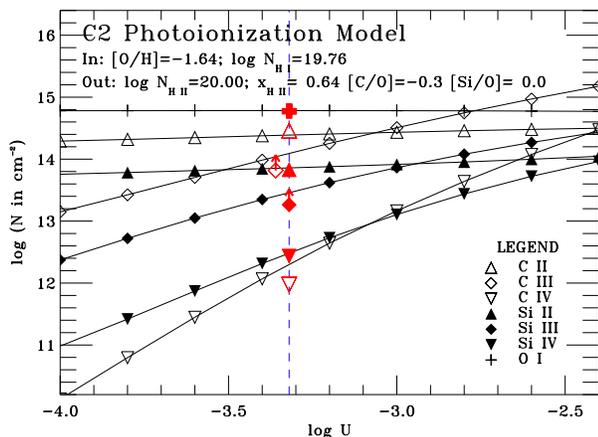} 
\caption{CLOUDY photoionization modelling of component C2, assuming
  the radiation field has contributions from both the quasar and the
  extragalactic background. Each series of connected symbols shows the
  predicted column density of a given ion as a function of ionization
  parameter (or equivalently, gas density, which increases to the
  left). The observations are shown in red, plotted at the density
  of the best-fit model.} 
\end{figure}

The ionization level we derive in the low-ion phase (64\%) 
is no larger than the levels derived in intervening sub-DLAs by other authors 
\citep{Pr99, Pe07}, reinforcing the idea that this sub-DLA does not
show unusual properties because of its proximity to the quasar.
The idea that many proximate DLAs (and by extension
sub-DLAs) may be representative of the intervening population has been
proposed before \citep[e.g.][]{Mo98}, though see \citet{Pr07}.

\section{A Highly Ionized Plasma Phase}
The \os\ and \cf\ profiles are clearly different from the low-ion
profiles, with the strongest high-ion component (H1) separated by 
100\,\kms\ from the strongest low-ion component.
\os\ and \cf\ show similar profiles at $-$150 to $-$100\,\kms, but
the \os\ absorption is smoother and broader than the
\cf\ in the range $-$100 to 0\,\kms. This can be seen in Figure 3,
where we compare the apparent column density profiles of
the two ions.
The difference in the profiles implies that the \os\ and \cf\ are not
fully co-spatial, i.e. the two ions trace different temperature
regions of a multi-phase structure\footnote{Similar differences 
  between the profiles of \cf\ and \os\ have been found before in many
  other environments, including the Milky Way halo \citep{Fo03},
  the Large Magellanic Cloud \citep{LH07}, LLSs \citep{KT97, KT99},
  DLAs \citep{Fo07a}, and some intergalactic absorbers \citep{Si02}.}.
We derive a lower limit on $N_{\rm H\,II}$ in the \os-bearing gas
using $N_{\rm Hot~H\,II}=N_{\rm O\,VI}/(f_{\rm O\,VI}{\rm O/H})$, 
with an \os\ ionization fraction $f_{\rm O\,VI}=0.2$ 
\citep[the maximum allowed by collisional ionization models; see the
  Appendix in][]{TS00} 
and assuming that O/H in the high-ionization gas is the same
as O/H in the low-ionization gas. This method yields 
log\,$N_{\rm Hot\,H\,II}>$20.09 (total over H1 and H2).

\begin{figure}
\epsscale{1.2}
\plotone{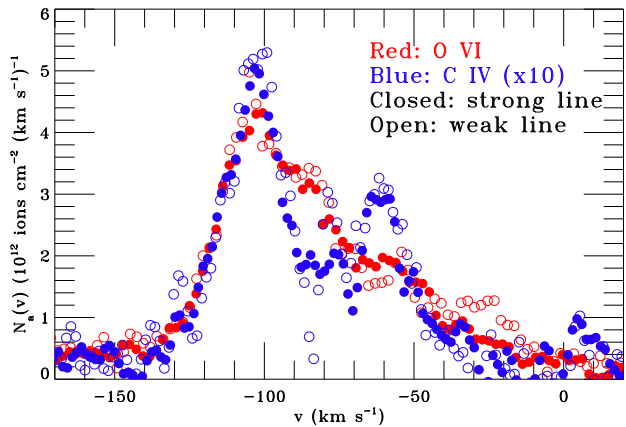}
\caption{Comparison of apparent column density profiles of \cf\ and
  \os\ in the sub-DLA. Both ions show similar profiles at $-$150 to
  $-$100\,\kms, but the \os\ absorption is smoother and broader than the
  \cf\ in the range $-$100 to 0\,\kms.}
\end{figure}

To explore how the \os\ properties of our sub-DLA relate to other systems,
we investigated the only intervening sub-DLA with a published \os\
detection: the system at $z_{\rm abs}=2.439$ toward
\object{Q1442+2931} \citep{Si02}, and the previously unpublished
\os\ profile in the sub-DLA at $z_{\rm abs}$=2.139 toward \object{Q1037-270}
\citep{Sr01}. The three sub-DLAs are described in Table 2.
Following the same calculation as used here,
the ratio $N_{\rm Hot H\,II}/N_{\rm H\,I}$ takes the value
$>$12 in the \object{Q1442+2931} sub-DLA, and
$>$1.0 in the \object{Q1037-270} sub-DLA,
vs $>$1.9 for the sub-DLA in this paper. 
From this we conclude that 
(i) in sub-DLAs, there are more baryons in the \os\ phase than there
are in the form of \hi\ atoms, and 
(ii) there is no evidence that radiation from the quasar is creating
more \os\ in the \object{Q0331-4505} sub-DLA than in intervening
sub-DLAs. 

\section{Contribution of Sub-DLAs to $\Omega$}
The latest determination of the cosmic density in \hi\ atoms in
sub-DLAs is $\Omega_{\rm H\,I}^{\rm sub\,DLA}$=$1.5\times10^{-4}$
\citep{OM07}. 
In order to fully account for the contribution of ionized gas in
sub-DLAs, it is necessary to determine the average ionized-to-neutral ratio
$\langle N_{\rm Total\,H\,II}/N_{\rm H\,I}\rangle$, where 
$N_{\rm Total\,H\,II}=N_{\rm Warm\,H\,II}+N_{\rm Hot\,H\,II}$. 
The value of $N_{\rm Warm\,H\,II}/N_{\rm H\,I}$ has been
investigated by \citet{Pe07}, who report a mean ionization fraction of
0.68 in 26 sub-DLAs modelled with CLOUDY, corresponding to 
$\langle N_{\rm Warm\,H\,II}/N_{\rm H\,I} \rangle$=2.1. 
The value of the $N_{\rm Hot\,H\,II}/N_{\rm H\,I}$ is not well
constrained: as discussed in \S5, the derived values in the three
existing sub-DLAs with \os\ are $>$1.0, $>$1.9 and $>$12. 
Assuming that the mean $N_{\rm Hot\,H\,II}/N_{\rm H\,I}$ in sub-DLAs
lies somewhere between these measurements, we find that 
$\langle N_{\rm Total\,H\,II}/N_{\rm H\,I} \rangle$ is between 3.1 and 14. 
Thus, the total baryon content of sub-DLAs
$\Omega_{\rm H\,I+H\,II}^{\rm sub\,DLA}=
(1+\langle N_{\rm Total\,H\,II}/N_{\rm H\,I} \rangle)
\Omega_{\rm H\,I}^{\rm sub\,DLA}$ is between 
$6.1\times10^{-4}$ and $2.2\times10^{-3}$. 

The total contribution of sub-DLAs to the metal budget then follows
according to:  
$\Omega_{Z,{\rm H\,I+H\,II}}^{\rm sub\,DLA}=\langle{\rm Z/H}\rangle
\Omega_{\rm H\,I+H\,II}^{\rm sub\,DLA}$. 
Assuming a typical sub-DLA metallicity of 
$\langle[{\rm Z/H}]\rangle$=$-$0.8 \citep{Ku07},
then $\Omega_{Z,{\rm H\,I+H\,II}}^{\rm sub\,DLA}$
is between $1.8\times10^{-6}$ and $6.7\times10^{-6}$,  
i.e between 6\% and 22\% of all the metals produced by star formation up to
$z\approx2$ \citep*[using a value for the total $\Omega_Z$ from star
formation of $3.0\times10^{-5}$;][]{Fe05}. 
For comparison, $\Omega_{Z,{\rm H\,I+H\,II}}^{\rm DLA}\approx6\times10^{-7}$
\citep*[including both the neutral and highly ionized phases;][]{Pr05, Fo07a}. 
This calculation shows that globally there are at least three times
more metal atoms in sub-DLAs than DLAs, owing to both the higher mean
metallicity and the higher mean ionization level in sub-DLAs. 

There are several reasons why this estimate of $\Omega_{Z}$ in sub-DLAs
could rise. First, recent detections of sub-DLAs
with super-solar metallicity \citep{Kh04, Pr06, Pe06} may imply
the mean sub-DLA metallicity is higher than $-$0.8.
Second, the typical temperature of the \os-bearing gas in sub-DLAs
may be above $3\!\times10^5$\,K, implying the \os\ ionization fraction will be
lower than 0.2 \citep{SD93}.
Finally, if the \os-bearing gas was generated by the accretion and
shock-heating of intergalactic gas, the ionized gas may have lower
metallicity than the neutral gas (increasing $N_{\rm Hot\,H\,II}$).
Each of these effects would serve to increase $\Omega_{Z}$ above the 6--22\%
level discussed here. 

\acknowledgments
AJF gratefully acknowledges support from a Marie Curie Intra-European
Fellowship awarded by the European Union Sixth Framework Programme.
We thank Bart Wakker for help in implementing CLOUDY, and Robert Simcoe
for clarifying the properties of the Q1442+2931 sub-DLA.

\clearpage
\input{tab2.tex}

\end{document}

%% file: tab1.tex
\begin{deluxetable}{llll}
\tablewidth{0pt}
\tabcolsep=2pt
\tablecaption{The Sub-DLA toward Q0331-4505}  
\tablehead{Property & Value & Property & Value}
\startdata 
$z_{\rm abs}$ & 2.65618 &
$z_{\rm qso}$                    & $2.6785\!\pm\!0.0030$\tm{a}\\

$d$                              & 7.1\,Mpc\tm{b}&
$v_{\rm qso}$--$v_{\rm abs}$     & $1\,820\!\pm\!250$\,\kms\\

log\,$N_{\rm H\,I}$              & $19.82\!\pm\!0.05$ &
log\,$N_{\rm O\,I}$              & $14.90\!\pm\!0.05$ \\

$\Delta v_{\rm Si\,II}$\tm{c}    & $76\!\pm\!2$\,\kms  &
$\Delta v_{\rm O\,VI}$\tm{c}     & $104\!\pm\!2$\,\kms \\

log\,$N_{\rm O\,VI}$             & $14.43\!\pm\!0.02$ &
log\,$N_{\rm C\,IV}$             & $13.37\!\pm\!0.02$ \\

log\,$N_{\rm Warm\,H\,II}$\tm{d} & $20.06\!\pm\!0.10$ &
${\rm [O/H]}$                    & $-1.64\!\pm\!0.07$\\

log\,$N_{\rm Hot\,H\,II}$\tm{e}   & $>$20.09 &
$N_{\rm Hot\,H\,II}$/$N_{\rm H\,I}$ & $>$1.9\\

log\,$N_{\rm Total\,H\,II}$      & $>$20.38 &
$N_{\rm Tot\,H\,II}$/$N_{\rm H\,I}$ & $>$3.6

\enddata
\tn{a}{$z_{\rm qso}$ is an average of all the QSO emission lines measured.\\}
\tn{b}{Estimated distance from the QSO, assuming Hubble flow (see text).\\}
\tn{c}{Velocity width containing 90\% of the integrated optical depth.\\}
\tn{d}{$N_{\rm Warm\,H\,II}$ is derived from CLOUDY
  photoionization modelling to the low-ion column densities (see text).\\} 
\tn{e}{The lower limits to $N_{\rm Hot\,H\,II}$ are
  found using the maximum allowed \os\ ionization fraction 
  ($f_{\rm O\,VI}$=0.2), and then calculating
  $N_{\rm H\,II}=N_{\rm O\,VI}/(f_{\rm O\,VI}{\rm O/H})$.
}
\end{deluxetable}

%% file: tab2.tex
\begin{deluxetable}{lllll llll}
\tablewidth{0pt}
\tabcolsep=2pt
\tablecaption{Three Sub-DLAs with \os\ Detections}  
\tablehead{QSO & $z_{\rm em}$ & $z_{\rm abs}$ & Type & log\,$N_{\rm H\,I}$ & [O/H] 
& log\,$N_{\rm O\,VI}$ & $N_{\rm Hot\,H\,II}/N_{\rm H\,I}$ & Ref.}
\startdata 
Q0331-4505 & 2.674 & 2.656 & Proximate   & 19.82$\pm$0.05 & $-$1.64$\pm$0.07 & 14.43$\pm$0.02 & $>$1.9  &1\\
Q1442+2931 & 2.669 & 2.439 & Intervening & 19.78$\pm$0.30 & $-$1.82$\pm$0.40 & 15.00$\pm$0.01 & $>$12.0 &2\\
Q1037-270  & 2.201 & 2.139 & Intervening & 19.70$\pm$0.05 & $-$0.31$\pm$0.05 & $\approx$15.35$\pm$0.30   & $>$1.0  &3
\enddata
\tablerefs{(1) this paper; (2) \citet{Si02}; (3) \citet{Sr01}.}
\end{deluxetable}